\documentclass{ifacconf}

\counterwithin*{section}{part}

\usepackage{graphicx}      
\usepackage{natbib}        
\AtEndEnvironment{pf}{~\hfill~$\qed$\par} 
\usepackage{amsmath}
\usepackage{mathtools}					
\usepackage{amssymb}					
\usepackage{bm}					
\usepackage[
	capitalise		
]{cleveref}
\crefname{section}{Sec.}{Sec.}
\Crefname{section}{Section}{Sections}
\crefname{figure}{Fig.}{Fig.}
\Crefname{figure}{Figure}{Figures}
\crefname{equation}{}{}
\Crefname{equation}{Equation}{Equations}
\crefrangeformat{equation}{(#3#1#4)\,--\,(#5#2#6)}		
\usepackage{autonum}
\newcommand{\retainlabel}[1]{\label{#1}\sbox0{\ref{#1}}} 
\usepackage{csquotes}
\usepackage{siunitx}
\DeclareSIUnit\bar{bar}
\usepackage{tikz}
\usepackage{nicematrix}
\NiceMatrixOptions{
	custom-line = { letter = I , tikz = dashed , total-width = \pgflinewidth }
}
\newcommand{\ie}{i.e.}
\newcommand{\cf}{cf.}
\newcommand{\eg}{e.g.}
\DeclarePairedDelimiter{\br}{[}{]}			
\DeclarePairedDelimiterX{\norm}[1]{\lVert}{\rVert}{#1}
\DeclarePairedDelimiterX{\abs}[1]{\lvert}{\rvert}{\ifblank{#1}{{}\cdot{}}{#1}}
\DeclareMathOperator*{\argmin}{\mathrm{argmin}}
\DeclarePairedDelimiterX{\closedinterval}[2]{[}{]}{#1,#2}
\DeclareMathOperator{\sign}{sign}
\DeclareMathOperator{\diag}{diag}
\newcommand{\set}[1]{\ensuremath{\left\{#1\right\}}}
\makeatletter
\def\@opargbegintheorem#1#2#3{\trivlist
   \item[]{\bfseries #1\ #2\ (#3)} \itshape}
\makeatother

\theoremheaderfont{\bfseries}

\newtheorem{lem_eigen}{Lemma}
\newtheorem{thm_eigen}{Theorem}
\newtheorem{rem_eigen}{Remark}

\usepackage{hyphenat}

\usepackage{subcaption}		

\usepackage{romannum}

\newcommand{\setSloppyPenalty}{\relpenalty=500}
\newcommand{\setHardPenalty}{\relpenalty=10000}

\usepackage{microtype}

\usepackage{eso-pic}

\begin{document}
\begin{frontmatter}

\title{Parameter-interval estimation for cooperative reactive sputtering processes}

\author[First]{Fabian Schneider}
\author[First]{Christian Wölfel}

\address[First]{Chair of Automation, Ruhr University Bochum, Universitaetsstrasse 150, 44801 Bochum, Germany, \{fabian.schneider-v8b, christian.woelfel\}@rub.de}

\begin{abstract}                
	Reactive sputtering is a plasma-based technique to deposit a thin film on a substrate.
	This contribution presents a novel parameter-interval estimation method for a well-established model that describes the uncertain and nonlinear reactive sputtering process behaviour.
	Building on a proposed monotonicity-based model classification, the method guarantees that all parameter values within the parameter interval yield output trajectories and static characteristics consistent with the enclosure induced by the parameter interval.
	Correctness and practical applicability of the new method are demonstrated by an experimental validation, which also reveals inherent structural limitations of the well-established process model for state-estimation tasks.
\end{abstract}


\end{frontmatter}

\AddToShipoutPictureFG*{
	\AtPageLowerLeft{%
		\makebox[\paperwidth]{%
			\raisebox{6mm}[0pt][0pt]{%
				\footnotesize%
				\copyright {} 2026 the authors.
				This work has been accepted to IFAC for publication under a Creative Commons Licence CC-BY-NC-ND.
			}%
		}%
	}%
}

\section{Introduction}

\subsection{Uncertain nonlinear reactive sputtering processes}

\begin{figure}[b]
	\begin{center}
		\includegraphics{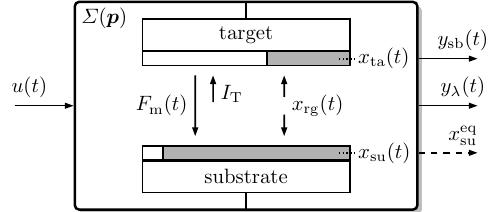}
		\caption{%
			Material flows in reactive sputtering described by the well-established process model~$\varSigma(\bm p)$~(\cf~\cref{eq:processModel:berg2014ss:system}).
		} 
		\label{fig:prozessBild}
	\end{center}
\end{figure}

Reactive sputtering is a plasma-based technique used to deposit thin films on a substrate, for example in the manufacturing of semiconductors and optical coatings.
The process executes in a vacuum chamber by means of a low-pressure plasma.
Material is sputtered from the target by an ion current~$I_\mathrm{T}$ (\cref{fig:prozessBild}).
The resulting metal flow~$F_\mathrm{m}(t)$ towards the substrate causes the thin film growth.
Due to a reactive gas in-flow~$u(t)$, a corresponding partial pressure~$x_\mathrm{rg}(t)$ builds up and leads to the formation of compound layers characterized by the surface fractions~$x_\mathrm{ta}(t)$ and~$x_\mathrm{su}(t)$.
It is a nonlinear process and the plasma-surface interaction is uncertain.

The state variables~$x_\mathrm{ta}(t)$,~$x_\mathrm{rg}(t)$ are indirectly given by~$y_\mathrm{sb}(t)$ and~$y_\lambda(t)$, which represent raw sensor signals.
Since~$x_\mathrm{su}(t)$ is an important film-quality indicator, reflecting the chemical composition, but
only its steady-state behaviour~$x_\mathrm{su}^\mathrm{eq}$ can be captured by ex-situ measurements, a monitoring problem arises.
This concerns the surface behaviour $x_\mathrm{su}(t)$ during process runtime.

An interval observer can be systematically designed based on the standard nonlinear process model~\citep{Berg:2005} to give guaranteed bounds on~$x_\mathrm{su}(t)$~\citep{Schneider:2023:ccta}.
However, the process model has in practice uncertain parameters and, therefore, guaranteed parameter intervals must be estimated to maintain the guaranteed bounds of the interval observer.

To address this estimation task, the present contribution introduces a novel method, based on a rigorous monotonicity analysis, to systematically determine parameter intervals for
the well-established process model~\citep{Berg:2014} of reactive sputtering.
Its practical applicability for state-estimation tasks is demonstrated by an experimental validation, which also reveals structural limitations of the well-established process model.

\subsection{Literature survey}

The standard model~\citep{Berg:2005} for reactive sputtering is based on flow-balance equations for metal, reactive gas and compound material.
Extensions~\citep{Kubart:2006,Berg:2014} of the standard model incorporate additional effects, such as dissociation of sputtered molecules and multiple layers and segments of the surfaces, but they increasingly complicate the design of the interval observer.

In the field of plasma and materials science, point values for the model parameters are identified by an interplay of different methods.
\cite{Kelemen:2021} identify parameters from analytically obtained properties of the static characteristics.
Specific parameters such as the sputtering yields can be determined by Monte Carlo simulations~\citep{Mahne:2022}.
\cite{StrijckmansLeroyGryseDepla:2012} present a method for estimating a parameter set guaranteeing a bounded model-output error.
These methods require substantial effort or yield parameter estimates of limited accuracy (or both).

Bounded-error estimation~\citep{Kieffer:2011,Jaulin:2001}, typically based on set inversion via interval analysis (SIVIA) and contractor techniques, aims to compute all boxes in the parameter space that are consistent with a user-defined error bound between the measurements and the model output.
In contrast, guaranteed global optimization~\citep{Kieffer:2011} determines parameter boxes containing the minimizers of a cost function involving the model output error.
Monotone dynamical systems,~\ie, systems that preserve the order of their states over time~\citep{Smith:1995}, facilitate parameter-interval estimation.
\cite{Angeli:2003} extend the monotonicity concept to controlled systems, as is the case in reactive sputtering.

This contribution introduces a new method for estimating two parameter vectors based on minimizing the model output error.
Instead of guaranteeing an enclosure for the parameter values under a user-defined error bound, the method guarantees that all parameter values within the resulting parameter interval lead to enclosed model outputs, which is necessary for interval-observer design.
A rigorous monotonicity analysis constitutes the basis for the parameter-interval estimation method.

\subsection{Structure of the contribution}

\Cref{sec:processModel,sec:problemFormulation} introduce the well-established process model and state the objectives~(\MakeUppercase{\romannumeral 1}) and~(\MakeUppercase{\romannumeral 2}) of the present work, respectively.
The rigorous monotonicity analysis in~\cref{sec:modelAnalysis:monotonicity} addresses objective~(\MakeUppercase{\romannumeral 1}), while \cref{sec:intervalEstimation} presents the new parameter-interval estimation method for objective~(\MakeUppercase{\romannumeral 2}).
\Cref{sec:validation,sec:conclusion} present the experimental validation results and summarize this contribution.

\section{State-space process model}
\label{sec:processModel}

The well-established first-principles process model~\citep{Berg:2014} is expressed in the state-space form
\begin{align}
	\label{eq:processModel:berg2014ss:system}
	\varSigma(\bm p)
	\colon
	\left\lbrace
	\begin{aligned}
		\dot{\bm x}(t)
		&
		=
		\bm f(\bm p, \bm x(t), u(t))
		, \quad
		\bm x(0)
		=
		\bm x_0
		\\
		\bm y(t)
		&=
		\bm g(\bm p, \bm x(t))
		.
	\end{aligned}
	\right.
\end{align}
The state~$
	\bm x(t) = \begin{pmatrix}
		x_\mathrm{rg}(t)
		&
		x_\mathrm{ta}(t)
		&
		x_\mathrm{su}(t)
	\end{pmatrix}^\top
$
and the measured output vector~$
	\bm y(t)
	=
	\begin{pmatrix}
		y_\lambda(t)
		&
		y_\mathrm{sb}(t)
	\end{pmatrix}^\top$
consist of the reactive gas partial pressure, the target surface coverage, the substrate surface coverage, and the measured raw sensor signals, respectively.
The system function~$\bm f$ and the output function~$\bm g$ are given as
\begin{subequations}
	\label{eq:processModel:berg2014ss:systemFunction}
\begin{alignat}{2}
	&f_1(\bm p, \bm x, u)
	&&=
	x_\mathrm{rg} \br{ - a_\mathrm{11} + a_\mathrm{112} (x_\mathrm{ta}-1) + a_\mathrm{113} (x_\mathrm{su}-1) }
	\\
	&&&\quad
	+ a_\mathrm{12}x_\mathrm{ta} - a_\mathrm{123} x_\mathrm{ta}(1-x_\mathrm{su}) + b u
	\retainlabel{eq:processModel:berg2014ss:systemFunction:xrg}
	\\
	&f_2(\bm p, \bm x)
	&&=
	a_\mathrm{212} x_\mathrm{rg} \br{1-x_\mathrm{ta}} - a_\mathrm{22} x_\mathrm{ta}
	\retainlabel{eq:processModel:berg2014ss:systemFunction:xta}
	\\
	&f_3(\bm p, \bm x)
	&&=
	a_\mathrm{313} x_\mathrm{rg} \br{1-x_\mathrm{su}} + a_\mathrm{323} x_\mathrm{ta} \br{1-x_\mathrm{su}} 
	\\
	&&&\quad
	- ( a_\mathrm{33c} x_\mathrm{ta} + a_\mathrm{33m} \br{1-x_\mathrm{ta}} ) x_\mathrm{su}
	\retainlabel{eq:processModel:berg2014ss:systemFunction:xsu}
\end{alignat}
\end{subequations}
and
\begin{subequations}
	\label{eq:processModel:berg2014ss:outputFunction}
\begin{align}
	y_\lambda
	&=
	g_1(\bm p, x_\mathrm{rg})
	=
	- c_\mathrm{11} \ln \left( c_\mathrm{12} x_\mathrm{rg} \right)
	\label{eq:processModel:berg2014ss:outputFunction:xrgyLambda}
	\\
	y_\mathrm{sb}
	&=
	g_2(\bm p, x_\mathrm{ta})
	=
	c_\mathrm{20} - c_\mathrm{21} x_\mathrm{ta}
	.
	\label{eq:processModel:berg2014ss:outputFunction:xtaysb}
\end{align}
\end{subequations}
Assembling the physically interpretable model parameters of~\cite{Berg:2014} into the parameter vector
\begin{align}
	\bm p =
	\begin{pmatrix}
		\begin{pNiceArray}{ccccc:cc}
			a_\mathrm{11} & a_\mathrm{112} & a_\mathrm{113} & a_\mathrm{12} & a_\mathrm{123} & a_\mathrm{212} & a_\mathrm{22}
		\end{pNiceArray}^\top
		\\
		\begin{pNiceArray}{cccc:c:cc:cc}
			a_\mathrm{313} & a_\mathrm{323} & a_\mathrm{33c} & a_\mathrm{33m} & b & c_\mathrm{1l} & c_\mathrm{12} & c_\mathrm{20} & c_\mathrm{21}
		\end{pNiceArray}^\top
	\end{pmatrix}
	,
	\label{eq:processModel:berg2014ss:parameterVector}
\end{align}
results in~\cref{eq:processModel:berg2014ss:parameterVector} having no direct physical interpretation.
However, constraints implied by the physical interpretation carry over to the new parameter space, which defines the permissible codomains~$
\mathbb{D}_\mathrm{x} = \{ \bm x \in \mathbb{R}_{\geq 0}^3 \mid x_\mathrm{rg} \leq x_\mathrm{rg,max}, x_\mathrm{ta} \leq 1, x_\mathrm{su} \leq 1 \}$,
$\mathbb{D}_\mathrm{u} = \mathbb{R}_{\geq 0}$,
$\mathbb{D}_\mathrm{p} = \{ \bm p \in \mathbb{R}_{\geq 0}^{16} \mid a_\mathrm{12} \geq a_\mathrm{123}, a_\mathrm{33m} \geq a_\mathrm{33c}, c_{12} x_\mathrm{rg,max} < 1 \}$,
$\mathbb{D}_\mathrm{y} = \{ \bm y \in \mathbb{R}_{\geq 0}^2 \mid \bm y = \bm g(\bm p, \bm x), \bm p \in \mathbb{D}_\mathrm{p}, \bm x \in \mathbb{D}_\mathrm{x} \}$,
for~$\bm x(t)$, $u(t)$, $\bm p$ and~$\bm y(t)$.
$\varSigma(\bm p)$, with~$\bm p$ being a physically constrained parameter vector, represents not a single model, but a class of systems determined by the admissible parameter values characterizing reactive sputtering.

Both the dynamic behaviour of~$\varSigma(\bm p)$ and the static characteristics of~$\varSigma(\bm p)$ are analyzed with regard to monotonicity and are used for the estimation method (\cref{sec:intervalEstimation}).
Solving~$\bm 0 = \bm f(\bm p, \bm x^\mathrm{eq}, u^\mathrm{eq})$ yields the static characteristics
\begin{subequations}
	\label{eq:processModel:static}
\begin{alignat}{2}
	u^\mathrm{eq}
	&=
	\pi_\mathrm{u}(\bm p, x^\mathrm{eq}_\mathrm{rg}, x^\mathrm{eq}_\mathrm{ta}, x^\mathrm{eq}_\mathrm{su})
	\label{eq:processModel:static:urg}
	\\
	&
	=
	- \frac{1}{b}
	(
		a_\mathrm{12} x^\mathrm{eq}_\mathrm{ta} - x^\mathrm{eq}_\mathrm{123} x^\mathrm{eq}_\mathrm{ta} (1-x^\mathrm{eq}_\mathrm{su})
		\\
		&\qquad
		+
		x^\mathrm{eq}_\mathrm{rg}
		\left(
			- a_\mathrm{11} + a_\mathrm{112} (x^\mathrm{eq}_\mathrm{ta}-1) + a_\mathrm{113} (x^\mathrm{eq}_\mathrm{su}-1) 
		\right)
	)
	\\
	x^\mathrm{eq}_\mathrm{su}
	&=
	\pi_\mathrm{su}(\bm p, x^\mathrm{eq}_\mathrm{rg}, x^\mathrm{eq}_\mathrm{ta})
	\label{eq:processModel:static:xsu}
	\\
	&=
	\frac{a_\mathrm{313} x^\mathrm{eq}_\mathrm{rg} + a_\mathrm{323} x^\mathrm{eq}_\mathrm{ta}}{a_\mathrm{313} x^\mathrm{eq}_\mathrm{rg} + a_\mathrm{323} x^\mathrm{eq}_\mathrm{ta} + a_\mathrm{33c} x^\mathrm{eq}_\mathrm{ta} + a_\mathrm{33m} (1-x^\mathrm{eq}_\mathrm{ta})}
	\\
	x^\mathrm{eq}_\mathrm{ta}
	&=
	\pi_\mathrm{ta}(\bm p, x^\mathrm{eq}_\mathrm{rg})
	=
	\frac{a_\mathrm{212} x^\mathrm{eq}_\mathrm{rg}}{a_\mathrm{212} x^\mathrm{eq}_\mathrm{rg} + a_\mathrm{22}}
	\label{eq:processModel:static:xta:basedOn:xrg}
	.
\end{alignat}
\end{subequations}

By composing functions~\cref{eq:processModel:static:urg,eq:processModel:static:xsu,eq:processModel:static:xta:basedOn:xrg}, the static characteristic
\begin{align}
	\bm \pi
	(\bm p, x^\mathrm{eq}_\mathrm{rg})
	=
	\hspace{-1pt}
	\begin{pmatrix*}[l]
		\pi_\mathrm{u\phantom{t}}(\bm p, x^\mathrm{eq}_\mathrm{rg}, \pi_2(\bm p, x^\mathrm{eq}_\mathrm{rg}), \pi_3(\bm p, x^\mathrm{eq}_\mathrm{rg}) )
		\\
		\pi_\mathrm{ta}(\bm p, x^\mathrm{eq}_\mathrm{rg})
		\\
		\pi_\mathrm{su}(\bm p, x^\mathrm{eq}_\mathrm{rg}, \pi_2(\bm p, x^\mathrm{eq}_\mathrm{rg}) )
	\end{pmatrix*}
	\hspace{-1pt}
	=
	\hspace{-1pt}
	\begin{pmatrix}
		u^\mathrm{eq}
		\\
		x^\mathrm{eq}_\mathrm{ta}
		\\
		x^\mathrm{eq}_\mathrm{su}
	\end{pmatrix}
	\label{eq:processModel:static:equilibriumPoint:ux:basedOn:xrg}
\end{align}
is determined for all~$x_\mathrm{rg}^\mathrm{eq} \in \mathbb{D}_\mathrm{x}$ and~$\bm p \in \mathbb{D}_\mathrm{p}$.
Based on~\cref{eq:processModel:static:equilibriumPoint:ux:basedOn:xrg} and the inverses
\begin{subequations}
	\label{eq:inverseOutputFunction}
\begin{alignat}{3}
	x_\mathrm{rg}
	&=
	g_1^{-1}(\bm p, y_\lambda)
	&&=
	\frac{1}{c_{12}} \mathrm{e}^{- \frac{y_\mathrm{\lambda}}{c_\mathrm{11}}}
	\label{eq:processModel:static:xrg:basedOn:yLambda}
	\\
	x_\mathrm{ta}
	&=
	g_2^{-1}(\bm p, y_\mathrm{sb})
	&&=
	\frac{c_\mathrm{20} - y_\mathrm{sb}}{c_\mathrm{21}}
	\label{eq:processModel:static:xta:basedOn:ysb}
\end{alignat}
\end{subequations}
of~\cref{eq:processModel:berg2014ss:outputFunction:xrgyLambda,eq:processModel:berg2014ss:outputFunction:xtaysb}
with respect to the state variables, the measurable steady-state values
can be determined from the static characteristic
\begin{align}
	\bm \varphi
	(\bm p, y^\mathrm{eq}_\lambda)
	=
	\begin{pmatrix}
		u^\mathrm{eq}
		&
		x^\mathrm{eq}_\mathrm{su}
		&
		y^\mathrm{eq}_\mathrm{sb}
	\end{pmatrix}^\top
	\label{eq:processModel:static:equilibriumPoint}
\end{align}
of~$\varSigma(\bm p)$
for all~$y^\mathrm{eq}_\lambda \in\mathbb{D}_\mathrm{y}$ and~$\bm p \in\mathbb{D}_\mathrm{p}$.
\cite{Berg:2005} show the typical S-shaped (non-monotone) static characteristic of~$\varSigma(\bm p)$.

\section{Problem formulation}
\label{sec:problemFormulation}

\subsection{Notation and preliminaries on monotonicity}
\label{sec:notation}

\setHardPenalty{}
$\bm 1$ denotes a column vector of ones.
For a vector $\bm x$, $\diag(\bm x)$ denotes the diagonal matrix with the entries of $\bm x$ on its diagonal.
Superscripts \enquote{a} and \enquote{b} mark parameters and signals that form corresponding counterparts (see \cref{fig:blockschaltbild_cooperativeSystem,fig:blockschaltbild_parameterMonotoneSystem}).
The sign function~$\sign(f(\cdot))$ evaluates if~$f(\cdot)$ is nonnegative ($+$), nonpositive ($-$) or zero for all considered arguments~$(\cdot)$.
Inequalities such as $\leq$ and $\geq$ are interpreted element-wise when applied to vectors.
A partial order relation, with possibly mixed monotonicity directions, is defined as~$
	\bm x^{\mathrm{a}} \preceq_{\bm r} \bm x^{\mathrm{b}}
	\iff
	\diag(\bm r)\,(\bm x^{\mathrm{a}} - \bm x^{\mathrm{b}}) \leq \bm 0
$
by a vector~$\bm r$ with entries $r_i \in \{+,0,-\}$.
For example,~$
	x^{\mathrm{a}}_1 \leq x^{\mathrm{b}}_1
	,
	x^{\mathrm{a}}_2 \geq x^{\mathrm{b}}_2
	\iff
	\bm x^{\mathrm{a}} \preceq_{\bm r} \bm x^{\mathrm{b}},
	\bm r = \begin{pmatrix} + & - \end{pmatrix}^\top
$.
A mapping $\bm y = \bm f(\bm x)$ is called \emph{monotone with respect to} $\preceq_{\bm r}$ if~$
	\bm x^{\mathrm{a}} \preceq_{\bm r} \bm x^{\mathrm{b}}
$
implies~$
	\bm y^{\mathrm{a}} \leq \bm y^{\mathrm{b}}
	.
$
\setSloppyPenalty{}
\begin{lem_eigen}[Monotone function]
\label{lem:monotoneViaGradient}
\phantom{x}
\\
\noindent
Consider a continuously differentiable function
$f \colon \mathbb{D}_{\mathrm{x}} \subseteq \mathbb{R}^n \to \mathbb{R}$.  
If~$
	\sign
	\left(
	\frac{\partial f}{\partial \bm x}(\bm x)
	\right)
	= \bm r
	\quad \forall\, \bm x \in \mathbb{D}_{\mathrm{x}},
$
then $f$ is monotone with respect to $\preceq_{\bm r}$,~\cf~\citep[p. 108]{Rudin:1976}.
\end{lem_eigen}

\subsection{Objectives}
\label{sec:performanceSpecifications}

(\MakeUppercase{\romannumeral 1})
An analysis is performed (\cref{sec:modelAnalysis:monotonicity}) to determine partial order relations~$\preceq_{\bm r_\mathrm{p}}$, $\preceq_{\bm r_\mathrm{x}}$, $\preceq_{r_\mathrm{u}}$, $\preceq_{\bm r_\mathrm{y}}$,~$\preceq_{\bm r_\varphi}$, such that ordered parameters,
initial conditions and manipulated signals
\begin{alignat}{5}
	&\bm p^\mathrm{a}
	&&\preceq_{\bm r_\mathrm{p}}
	&&\bm p^\mathrm{b}
	\label{eq:aim:order:parameter}
\\
	&\bm x_0^\mathrm{a}
	&&\preceq_{\bm r_\mathrm{x}}
	&&\bm x_0^\mathrm{b}
	\label{eq:aim:order:initialCondition}
\\
	\forall t \geq 0
	\colon
	&u^\mathrm{a}(t)
	&&\preceq_{r_\mathrm{u}}
	&&u^\mathrm{b}(t)
	\label{eq:aim:order:manipulatedSignal}
\end{alignat}
out of their respective domains~$\mathbb{D}_\mathrm{p}$~$\mathbb{D}_\mathrm{x}$ and~$\mathbb{D}_\mathrm{u}$ lead to ordered corresponding state trajectories
\begin{align}
	\forall t \geq 0
	\colon
	\bm y^\mathrm{a}(t)
	\preceq_{\bm r_\mathrm{y}}
	\bm y^\mathrm{b}(t)
	.
	\label{eq:performanceSpecification:enclose:trajectory}
\end{align}
Furthermore,~\cref{eq:aim:order:parameter,eq:aim:order:initialCondition,eq:aim:order:manipulatedSignal} in conjunction with (\cf~Remark~\labelcref{rem:why_b_needsToBeEqual})
\begin{align}
	c_{11}^\mathrm{a} = c_{11}^\mathrm{b}
	, \qquad
	c_{12}^\mathrm{a} = c_{12}^\mathrm{b}
	\label{eq:performanceSpecification:restriction:c11c12}
\end{align}
must ensure
ordered corresponding static characteristics
\begin{align}
	\forall y^\mathrm{eq}_\lambda \in \mathbb{D}_\mathrm{y}
	\colon
	\bm \varphi(\bm p^\mathrm{a}, y^\mathrm{eq}_\lambda)
	\preceq_{\bm r_\varphi}
	\bm \varphi(\bm p^\mathrm{b}, y^\mathrm{eq}_\lambda)
	.
	\label{eq:performanceSpecification:enclose:static}
\end{align}

(\MakeUppercase{\romannumeral 2})
The objective of the new parameter-interval estimation method (\cref{sec:intervalEstimation}) is to estimate two non-trivial parameter vectors, satisfying~\cref{eq:aim:order:parameter}, such that the measured trajectory~$\bm y^\mathrm{M}(t)$ and the measured steady-state values~$\bm \varphi^\mathrm{M}(y^\mathrm{eq}_\lambda)$ are enclosed:
\begin{alignat}{5}
	\forall t \geq 0
	&\colon
	\bm y^\mathrm{a}(t)
	&&\preceq_{\bm r_\mathrm{y}}
	\bm y^\mathrm{M}(t)
	&&\preceq_{\bm r_\mathrm{y}}
	\bm y^\mathrm{b}(t)
	\label{eq:performanceSpecification:encloseMeasurement:trajectory}
	\\
	\forall y^\mathrm{eq}_\lambda \in \mathbb{D}_\mathrm{y}
	&\colon
	\bm \varphi(\bm p^\mathrm{a}, y^\mathrm{eq}_\lambda)
	&&\preceq_{\bm r_\varphi}
	\bm \varphi^\mathrm{M}(y^\mathrm{eq}_\lambda)
	&&\preceq_{\bm r_\varphi}
	\bm \varphi(\bm p^\mathrm{b}, y^\mathrm{eq}_\lambda)
	.
	\label{eq:performanceSpecification:encloseMeasurement:static}
\end{alignat}

\section{Monotonicity analysis}
\label{sec:modelAnalysis:monotonicity}

\subsection{Section overview}
\label{sec:modelAnalysis:overview}

\begin{figure}
	\centering
	\begin{subfigure}[t]{1\linewidth}
		\centering
		\includegraphics{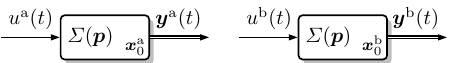}
		\caption{%
			Ordered initial conditions in~\cref{eq:aim:order:initialCondition},
			ordered manipulated signals~\cref{eq:aim:order:manipulatedSignal},
			with the same parameter vector~$\bm p$.
		}
		\label{fig:blockschaltbild_cooperativeSystem}
	\end{subfigure}
	\\
	\begin{subfigure}[t]{1\linewidth}
		\centering
		\includegraphics{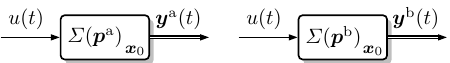}
		\caption{%
			Ordered parameter vectors in~\cref{eq:aim:order:parameter},
			with the same initial condition~$\bm x_0$ and the same input signal~$u(t)$.
		}
		\label{fig:blockschaltbild_parameterMonotoneSystem}
	\end{subfigure}
	\caption{%
		Ordered quantities of a (parameter-) monotone system lead to ordered output trajectories~\cref{eq:performanceSpecification:enclose:trajectory}.
	}
	\label{fig:blockschaltbild_cooperative_and_parameterMonotone_systems}
\end{figure}

This section addresses objective~(\MakeUppercase{\romannumeral 1}) by classifying~$\varSigma(\bm p)$ with respect to its monotonicity---both for fixed parameters~(\cref{sec:modelAnalysis:monotonicity:fixedParameters}, \cref{fig:blockschaltbild_cooperativeSystem}) and with respect to parameter variations~(\cref{sec:modelAnalysis:parameterMonotonicity}, \cref{fig:blockschaltbild_parameterMonotoneSystem}).
The partial order relations in~\cref{%
	eq:aim:order:parameter,%
	eq:aim:order:initialCondition,%
	eq:aim:order:manipulatedSignal,%
	eq:performanceSpecification:restriction:c11c12,%
	eq:performanceSpecification:enclose:trajectory,%
	eq:performanceSpecification:enclose:static%
} are chosen to align increases in parameter values (in the sense of~$\preceq_{\bm r_\mathrm{p}}$) with the model's inherent monotonicity direction ($\preceq_{\bm r_\mathrm{x}}$, $\preceq_{r_\mathrm{u}}$, $\preceq_{\bm r_\mathrm{y}}$, $\preceq_{\bm r_\varphi}$).

\subsection{Monotonicity under fixed parameter values}
\label{sec:modelAnalysis:monotonicity:fixedParameters}

First, the monotonicity of~$\varSigma(\bm p)$ under fixed parameter values is analyzed,~\ie, the implication from~\cref{%
	eq:aim:order:initialCondition,%
	eq:aim:order:manipulatedSignal,%
} to \cref{eq:performanceSpecification:enclose:trajectory,eq:performanceSpecification:enclose:static}.

\begin{lem_eigen}[Monotone output function]
	\label{lem:processModell:monotoneOutputFunction:xy}
	\phantom{x}
	\\
	\noindent
	For all~$\bm p \in \mathbb{D}_\mathrm{p}$ and all vectors~$\bm x_{12}^\mathrm{a}, \bm x_{12}^\mathrm{b} \in \mathbb{D}_\mathrm{x}$ of the first two state variables that correspond to~$\bm y^\mathrm{a}, \bm y^\mathrm{b} \in \mathbb{D}_\mathrm{y}$, the output function~\cref{eq:processModel:berg2014ss:outputFunction} and its inverse~\cref{eq:inverseOutputFunction} are monotone decreasing,~\ie
	\begin{align}
		\bm x_{12}^\mathrm{a} \leq \bm x_{12}^\mathrm{b}
		\iff
		\bm y^\mathrm{a}
		&\preceq_{\bm r_\mathrm{y}}
		\bm y^\mathrm{b}
		, \quad
		\label{eq:lem:processModell:monotoneOutputFunction:xy}
		\\
		\text{with }
		\bm r_\mathrm{y}
		&=
		\begin{pmatrix}
			- & -
		\end{pmatrix}^\top
		.
		\label{eq:relationOrderVector:y}
	\end{align}
\end{lem_eigen}
\begin{pf}
	Summarize~\cref{%
		eq:processModel:static:xrg:basedOn:yLambda,%
		eq:processModel:static:xta:basedOn:ysb,%
		eq:processModel:berg2014ss:outputFunction:xrgyLambda,%
		eq:processModel:berg2014ss:outputFunction:xtaysb%
	} by~$\bm g_{12}$ and~$\bm g_{12}^{-1}$.
	The equivalence in~\cref{eq:lem:processModell:monotoneOutputFunction:xy} is verified by determining the sign pattern~$
		\begin{pmatrix}
			- & 0
			\\
			0 & -
		\end{pmatrix}
		=
		\sign
		\frac{
			\partial
		}{
			\partial 
			\bm x_{12}
		}
		\bm g_{12}(\bm p, \bm x_{12})
		=
		\sign
		\frac{
			\partial
		}{
			\partial 
			\bm y
		}
		\bm g_{12}^{-1}(\bm p, \bm y)
	$
	of the gradients of~\cref{eq:processModel:berg2014ss:outputFunction} and of~\cref{eq:inverseOutputFunction} and by applying Lemma~\labelcref{lem:monotoneViaGradient} row-wise.
\end{pf}

Lemma~\labelcref{lem:processModell:monotoneOutputFunction:xy} addresses the monotonicity of an algebraic equation.
In contrast, the dynamical system~\cref{eq:processModel:berg2014ss:system} is called \textit{monotone}~\citep{Smith:1995,Angeli:2003} if a partial order of the state variables is preserved over time:
\begin{align}
	\forall
	t \geq 0
	\colon
	\bm x^\mathrm{a}_0
	\preceq_{\bm r_\mathrm{x}}
	\bm x^\mathrm{b}_0
	, \ 
	u^\mathrm{a}(t)
	\preceq_{r_\mathrm{u}}
	u^\mathrm{b}(t)
	\implies
	\bm x^\mathrm{a}(t)
	\preceq_{\bm r_\mathrm{x}}
	\bm x^\mathrm{b}(t)
	.
	\label{eq:def:monotoneSystem}
\end{align}

A \textit{cooperative} system is a special case of a monotone system~\citep{Angeli:2003} for
\begin{align}
	\bm r_\mathrm{x} = + \bm 1
	, \quad
	r_\mathrm{u} = + 1
	.
	\label{eq:relationOrderVector:xu}
\end{align}

\begin{figure}[t]
	\centerline{%
		\includegraphics{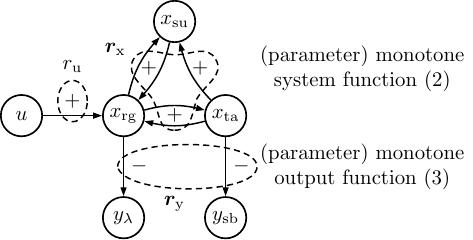}
		}
		\caption{%
			Structure graph of~$\varSigma(\bm p)$ indicating monotonically increasing (+) and monotonically decreasing (-) couplings between~$u(t)$ and~$\bm x(t)$ (Lemmas~\labelcref{lem:processModell:parameterMonotone,lem:processModell:cooperative}) and between~$\bm x(t)$ and~$\bm y(t)$ (Lemmas~\labelcref{lem:processModell:monotone:p:to:x:y,lem:processModell:monotoneOutputFunction:xy}).
		}
	\label{fig:strukturgraph_cooperativeSysFunc_monotoneOutFunc}
\end{figure}

\begin{lem_eigen}[Cooperative class of systems~$\varSigma(\bm p)$]
	\label{lem:processModell:cooperative}
	\hfill
	\\
	\noindent
	Reactive sputtering processes described by the class of systems $\varSigma(\bm p)$ are cooperative for all~$\bm p \in \mathbb{D}_\mathrm{p}$,
	$\bm x_0 \in \mathbb{D}_\mathrm{x}$,
	$u(t) \in \mathbb{D}_\mathrm{u}, t \geq 0$.
\end{lem_eigen}
\begin{pf}
	$
		\sign
		\frac{
			\partial
		}{
			\partial
			\bm x
		}
		\bm f(\bm p, \bm x, u)
		\!
		=
		\!
		\begin{pmatrix}
			- & + & +
			\\
			+ & - & +
			\\
			+ & + & -
		\end{pmatrix}
	$,
	$
		\sign
		\frac{
			\partial
		}{
			\partial
			u
		}
		\bm f(\bm p, \bm x, u)
		\!
		=
		\begin{pmatrix}
			+
			&
			0
			&
			0
		\end{pmatrix}^\top
	$
	are constant 
	$\forall \bm p \in \mathbb{D}_\mathrm{p}$ and satisfy Proposition~III.2 in~\cite{Angeli:2003}, from which it follows that~$\varSigma(\bm p)$
	is cooperative.
\end{pf}

The partial order relations~\cref{eq:relationOrderVector:y,eq:relationOrderVector:xu} (Lemmas~\labelcref{lem:processModell:monotoneOutputFunction:xy} and~\labelcref{lem:processModell:cooperative}, \cref{fig:strukturgraph_cooperativeSysFunc_monotoneOutFunc}) show that the state variables positively influence each other,
\eg,
increasing the reactive gas partial pressure and the target coverage increases the substrate coverage, and that the sensors are suitable as they yield a one-to-one correspondence with the first two state variables.

Cooperativity does not imply that the static characteristic~\cref{eq:processModel:static:equilibriumPoint:ux:basedOn:xrg} is monotone in~$x_\mathrm{rg}^\mathrm{eq}$ as well.

\begin{lem_eigen}[(Non-)Monotone static characteristic]
	\label{lem:analysis:monotone:staticCharacteristic}
	\phantom{x}
	\\
	\noindent
	The static characteristic~\cref{eq:processModel:static:equilibriumPoint:ux:basedOn:xrg} is monotone with respect to~$
		\widetilde{\bm r}_\pi
		=
		\begin{pmatrix}
			0 & + & +
		\end{pmatrix}^\top
	$
	for all~$\bm p \in \mathbb{D}_\mathrm{p}$ and~$x_\mathrm{rg}^\mathrm{eq} \in \mathbb{D}_\mathrm{x}$,~\ie,
	\begin{alignat}{3}
		x_\mathrm{rg}^\mathrm{eq,a}
		\leq
		x_\mathrm{rg}^\mathrm{eq,b}
		\implies
		\begin{pmatrix}
			x^\mathrm{eq,a}_\mathrm{ta}
			&
			x^\mathrm{eq,a}_\mathrm{su}
		\end{pmatrix}^\top
		\preceq_{\widetilde{\bm r}_\pi}
		\begin{pmatrix}
			x^\mathrm{eq,b}_\mathrm{ta}
			&
			x^\mathrm{eq,b}_\mathrm{su}
		\end{pmatrix}^\top
		\label{eq:analysis:monotone:static:pi:x}
		.
	\end{alignat}
	For some~$\bm p \in \mathbb{D}_\mathrm{p}$, the monotonicity of the first row of~\cref{eq:processModel:static:equilibriumPoint:ux:basedOn:xrg},~\ie,~$x_\mathrm{rg}^\mathrm{eq} \mapsto u^\mathrm{eq}$, is undefined.
\end{lem_eigen}
\begin{pf}
	The proof follows the same line of reasoning as the proof of Lemma~\labelcref{lem:processModell:monotoneOutputFunction:xy} by determining~$
		\sign
		\frac{\partial}{\partial x_\mathrm{rg}^\mathrm{eq}}
		\bm \pi(\bm p, x^\mathrm{eq}_\mathrm{rg})
		=
		\begin{pmatrix}
			*
			&
			+
			&
			+
		\end{pmatrix}^\top
	$.
	The mapping~$x_\mathrm{rg}^\mathrm{eq} \mapsto u^\mathrm{eq}$ is not monotone, since the sign of the first row is not constant (denoted by~$*$) for all~$x_\mathrm{rg}^\mathrm{eq} \in \mathbb{D}_\mathrm{x}$ and~$\bm p \in \mathbb{D}_\mathrm{p}$, which results from~$
		\sign
		\frac{\partial}{\partial \bm x}
		\pi_\mathrm{u}(\bm p, x^\mathrm{eq}_\mathrm{rg}, x^\mathrm{eq}_\mathrm{ta}, x^\mathrm{eq}_\mathrm{su})
		=
		\begin{pmatrix}
			+ & - & -
		\end{pmatrix}^\top
	$.
\end{pf}

The non-monotonicity of~$x_\mathrm{rg}^\mathrm{eq} \mapsto u^\mathrm{eq}$ (\cref{fig:fig_identResult_static}) occurs because increasing~$x_\mathrm{rg}^\mathrm{eq}$ increases the target coverage, reduces the sputtering rate and substrate reactive gas consumption and thus a smaller reactive gas inflow~$u^\mathrm{eq}$ is needed to maintain the same~$x_\mathrm{rg}^\mathrm{eq}$.

\begin{figure}[tb]
	\centerline{%
		\includegraphics{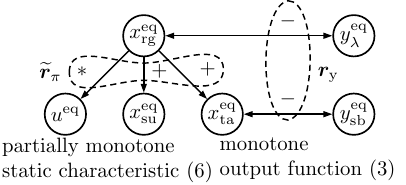}
		}
		\caption{%
			$\widetilde{\bm r}_{\bm \pi}$,~$\bm r_\mathrm{y}$ represent monotonically increasing (+) and monotonically decreasing (-) couplings in the structure graph of the static characteristic~\cref{eq:performanceSpecification:enclose:static}.
		}
	\label{fig:strukturgraph_monotoneStaticFunc_xrg_monotoneOutFunc}
\end{figure}

This section has shown that~$\varSigma(\bm p)$ and~\cref{eq:processModel:berg2014ss:outputFunction} are monotone in the sense of~\cref{%
	eq:aim:order:initialCondition,%
	eq:aim:order:manipulatedSignal,%
	eq:performanceSpecification:enclose:trajectory,%
} with the partial order relations~\cref{%
	eq:relationOrderVector:y,%
	eq:relationOrderVector:xu,%
}~(Lemmas~\labelcref{%
	lem:processModell:monotoneOutputFunction:xy,%
	lem:processModell:cooperative,%
})
under a fixed parameter vector~$\bm p \in \mathbb{D}_\mathrm{p}$ (\cref{fig:blockschaltbild_cooperativeSystem}).
The non-monotonicity of the static characteristic~$x_\mathrm{rg}^\mathrm{eq} \mapsto u^\mathrm{eq}$~(Lemma~\labelcref{lem:analysis:monotone:staticCharacteristic})
constitutes a structural limitation of the parameter-interval estimation method~(\cref{sec:intervalEstimation}) with respect to the number of parameter intervals that can be estimated (Remark~\labelcref{rem:why_b_needsToBeEqual}).
Nevertheless, the relation~$x_\mathrm{rg}^\mathrm{eq} \mapsto u^\mathrm{eq}$ is monotone with respect to parameter variations, as shown in the next section.

\subsection{Monotonicity with respect to parameter variations}
\label{sec:modelAnalysis:parameterMonotonicity}

In the following, the monotonicity of~$\varSigma(\bm p)$ with respect to the parameter vector~$\bm p$  (\cref{fig:blockschaltbild_parameterMonotoneSystem}) is analyzed.

\begin{lem_eigen}[Parameter-monotone output function]
	\label{lem:processModell:monotone:p:to:x:y}
	\phantom{x}
	\\
	\noindent
	Consider~\cref{eq:relationOrderVector:y},~$\bm p^\mathrm{a}, \bm p^\mathrm{b} \in \mathbb{D}_\mathrm{p}$ and
	\begin{align}
		\bm r^\mathrm{g}_\mathrm{p}
		&=
		\begin{pNiceArray}{ccc:cc:cc}
			0 & \hdots & 0 & - & + & - & +
		\end{pNiceArray}^\top
		.
		\label{eq:relatioOrderDefinitionMatrix:Rgp}
	\end{align}
	Then, the output function~\cref{eq:processModel:berg2014ss:outputFunction} and its inverse~\cref{eq:inverseOutputFunction} are monotone in the sense that
	\begin{alignat}{3}
		\forall \bm x_{12} \in \mathbb{D}_\mathrm{x}
		&\colon
		\bm p^\mathrm{a}
		\preceq_{\bm r^\mathrm{g}_\mathrm{p}}
		\bm p^\mathrm{b}
		\implies
		\bm y^\mathrm{a}
		&&\preceq_{\bm r_\mathrm{y}}
		&&\bm y^\mathrm{b}
		\label{eq:analysis:result:monotonicity:p:to:y}
	\\
		\forall \bm y \in \mathbb{D}_\mathrm{y}
		&\colon
		\bm p^\mathrm{a}
		\preceq_{\bm r^\mathrm{g}_\mathrm{p}}
		\bm p^\mathrm{b}
		\implies
		\bm x_{12}^\mathrm{a}
		&&\preceq_{\bm r_\mathrm{y}}
		&&\bm x_{12}^\mathrm{b}
		.
		\label{eq:analysis:result:monotonicity:p:to:x}
	\end{alignat}
\end{lem_eigen}
\begin{pf}
	For~$\bm p \in \mathbb{D}_\mathrm{p}$,
	$\bm x \in \mathbb{D}_\mathrm{x}$,
	$\bm y \in \mathbb{D}_\mathrm{y}$,
	the column sums of~$
		\sign
		\frac{\partial}{\partial \bm p} \bm g_{12}(\bm p, \bm x)
		=
		\sign
		\frac{\partial}{\partial \bm p} \bm g_{12}^{-1}(\bm p, \bm y)
		=
		\begin{pNiceArray}{ccc:cc:cc}
			0 & \hdots & 0 & + & - & 0 & 0
			\\
			0 & \hdots & 0 & 0 & 0 & + & -
		\end{pNiceArray}
	$
	are equal to
	$- \bm r^\mathrm{g\top}_\mathrm{p}$,
	which proves~\cref{eq:analysis:result:monotonicity:p:to:y,eq:analysis:result:monotonicity:p:to:x} according to Lemma~\labelcref{lem:monotoneViaGradient}.
\end{pf}

In contrast to dynamical-system monotonicity~(\cf~\cref{eq:def:monotoneSystem}), which considers different initial conditions and input signals~(\cref{fig:blockschaltbild_cooperativeSystem}), the following definition considers different parameter vectors while keeping the initial conditions and the input signal fixed~(\cref{fig:blockschaltbild_parameterMonotoneSystem}).

\begin{defn}[Parameter-monotone system]
	\label{def:parameterMonotoneSystem}
	\phantom{x}
	\\
	\noindent
	The system~\cref{eq:processModel:berg2014ss:system} is called \textit{parameter-monotone with respect to~$\preceq_{\bm r^\mathrm{f}_\mathrm{p}}$} if the implication
	\begin{align}
		\bm p^\mathrm{a}
		\preceq_{\bm r^\mathrm{f}_\mathrm{p}}
		\bm p^\mathrm{b}
		\implies
		\bm x^\mathrm{a}(t)
		\leq
		\bm x^\mathrm{b}(t)
		, \ 
		t \geq 0
		\label{eq:def:parameterMonotoneSystem}
	\end{align}
	holds for all~$\bm p^\mathrm{a}, \bm p^\mathrm{b}\in \mathbb{D}_\mathrm{p}$, $\bm x_0 \in \mathbb{D}_\mathrm{x}$ and all~$ t \geq 0 \colon u(t) \in \mathbb{D}_\mathrm{u}$.
\end{defn}

\begin{lem_eigen}[Parameter-monotone class of systems~$\varSigma(\bm p)$]
	\label{lem:processModell:parameterMonotone}
	Reactive sputtering processes described by the class of systems~$\varSigma(\bm p)$
	are parameter-monotone with respect to~$\preceq_{\bm r^\mathrm{f}_\mathrm{p}}$ with
	\begin{align}
		\bm r^\mathrm{f}_\mathrm{p}
		&=
		\begin{pNiceArray}{ccccc:cc:cccc:c:cc:cc}
			- & - & - & + & - &  + & - & + & + & - & - & + & 0 & 0 & 0 & 0
		\end{pNiceArray}^\top
		.
		\label{eq:relatioOrderDefinitionMatrix:Rfp}
	\end{align}
\end{lem_eigen}
\begin{pf}
	The system function~\cref{eq:processModel:berg2014ss:systemFunction} is monotone with respect to~$\preceq_{\bm r^\mathrm{f}_\mathrm{p}}$, since the column sums of~$
		\sign \frac{\partial}{\partial \bm p} \bm f(\bm p, \bm x, u)
		=
		\begin{pNiceArray}{ccccc:cc:cccc:c:cc:cc}
			- & - & - & + & - &   &   &   &   &   &   & + & 0 & 0 & 0 & 0
			\\
			  &   &   &   &   & + & - &   &   &   &   &   & 0 & 0 & 0 & 0
			\\
			  &   &   &   &   &   &   & + & + & - & - &   & 0 & 0 & 0 & 0
		\end{pNiceArray}
	$ are equal to $\bm r^\mathrm{f\top}_\mathrm{p}$
	for all~$\bm p \in \mathbb{D}_\mathrm{p}$, $\bm x \in \mathbb{D}_\mathrm{x}$, $u \in \mathbb{D}_\mathrm{u}$ due to~Lemma~\labelcref{lem:monotoneViaGradient}.
	Considering each parameter~$p_i$
	as a constant input signal and applying Corollary~III.3 from~\cite{Angeli:2003} proves~\cref{eq:def:parameterMonotoneSystem}.
\end{pf}

Lemmas~\labelcref{lem:processModell:monotone:p:to:x:y,lem:processModell:parameterMonotone} state that if the parameter vector~\cref{eq:processModel:berg2014ss:parameterVector}
is increased with respect to~$\preceq_{\bm r^\mathrm{f}_\mathrm{p}}$,
$\bm x(t)$ and~$\bm y(t)$ increase in the sense of~$\bm r_\mathrm{x}$ and~$\bm r_\mathrm{y}$
mo\-not\-o\-nous\-ly as shown in~\cref{fig:strukturgraph_cooperativeSysFunc_monotoneOutFunc},~\ie, to the same direction as described by Lemmas~\labelcref{lem:processModell:cooperative,lem:processModell:monotoneOutputFunction:xy}.

\begin{lem_eigen}[Parameter-monotone static characteristics]
	\label{lem:processModell:monotone:p:static}
	\hfill
	\\
	\noindent
	Consider two parameter vectors~$\bm p^\mathrm{a}, \bm p^\mathrm{b} \in \mathbb{D}_\mathrm{p}$.
	The static characteristic~\cref{eq:processModel:static:equilibriumPoint:ux:basedOn:xrg} is
	monotone such that for~\cref{eq:relatioOrderDefinitionMatrix:Rfp} and~$
		\bm r_{\pi}
		=
		\begin{pmatrix}
			- & + & +
		\end{pmatrix}^\top
	$ and all~$x_\mathrm{rg}^\mathrm{eq} \in \mathbb{D}_\mathrm{x}$
	the following implication holds:
	\begin{align}
		\bm p^\mathrm{a}
		&\preceq_{\bm r^\mathrm{f}_\mathrm{p}}
		\bm p^\mathrm{b}
		\\
		\implies
		\begin{pmatrix}
			u^\mathrm{a,eq}_\mathrm{rg}
			&
			x^\mathrm{a,eq}_\mathrm{ta}
			&
			x^\mathrm{a,eq}_\mathrm{su}
		\end{pmatrix}
		&\preceq_{\bm r_{\pi}}
		\begin{pmatrix}
			u^\mathrm{b,eq}_\mathrm{rg}
			&
			x^\mathrm{b,eq}_\mathrm{ta}
			&
			x^\mathrm{b,eq}_\mathrm{su}
		\end{pmatrix}
		.
		\label{eq:staticMonotoneIn:p}
	\end{align}
\end{lem_eigen}
\begin{pf}
	Analogously to the proofs of Lemmas~\labelcref{lem:processModell:monotone:p:to:x:y,lem:processModell:parameterMonotone}, Lemma~\labelcref{lem:processModell:monotone:p:static} is proven with
	$
		\sign
		\frac{\partial}{\partial \bm p}
		\bm \pi (\bm p, x^\mathrm{eq}_\mathrm{rg})
	$
	\\
	$
		=
		\begin{pNiceArray}{ccccc:cc:cccc:c:cccc}
			+ & + & + & - & + & - & + & - & - & + & + & - & 0 & 0 & 0 & 0
			\\
			0 & 0 & 0 & 0 & 0 & + & - & 0 & 0 & 0 & 0 & 0 & 0 & 0 & 0 & 0
			\\
			0 & 0 & 0 & 0 & 0 & + & - & + & + & - & - & 0 & 0 & 0 & 0 & 0
			\\
		\end{pNiceArray}
	$.
\end{pf}

\begin{figure}[tb]
	\centerline{%
		\includegraphics{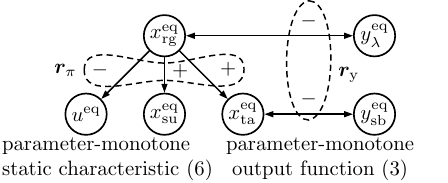}
		}
		\caption{%
			$\bm r_{\bm \pi}$ and~$\bm r_\mathrm{y}$ represent monotonically increasing~(+) and monotonically decreasing~(-) couplings in the structure graph of the static characteristic~\cref{eq:performanceSpecification:enclose:static}
			for an increase in~$\bm p$ in the direction~$\bm r_\mathrm{p}$.
		}
	\label{fig:strukturgraph_parameterMonotoneStaticFunc_OutFunc}
\end{figure}

Lemma~\labelcref{lem:processModell:monotone:p:static}, in conjunction with Lemma~\labelcref{lem:processModell:monotone:p:to:x:y}, states that increasing the parameter vector~$\bm p$ with respect to~$\preceq_{\bm r^\mathrm{f}_\mathrm{p}}$ leads to an increase in~$x^\mathrm{eq}_\mathrm{ta}$ and in~$x^\mathrm{eq}_\mathrm{su}$, and to a decrease in~$u^\mathrm{eq}$, in~$y^\mathrm{eq}_\lambda$ and in~$y^\mathrm{eq}_\mathrm{sb}$, as illustrated in~\cref{fig:strukturgraph_parameterMonotoneStaticFunc_OutFunc}.

\subsection{Main theorem}
\label{sec:modelAnalysis:boundingSystems}

Since no element of~$\bm p$ appears in the system function~\cref{eq:processModel:berg2014ss:systemFunction} and concurrently in the output function~\cref{eq:processModel:berg2014ss:outputFunction},~$\bm r^\mathrm{f}_\mathrm{p}$ ($\bm r^\mathrm{g}_\mathrm{p}$) defines no order for the parameters of the output function (system function).
Thus, the partial order relation
\begin{align}
	\bm r_\mathrm{p}
	=
	\bm r^\mathrm{f}_\mathrm{p}
	+
	\bm r^\mathrm{g}_\mathrm{p}
	\label{eq:relationOrderVector:p}
\end{align}
(\cf~\cref{eq:relatioOrderDefinitionMatrix:Rgp,eq:relatioOrderDefinitionMatrix:Rfp}) defines an order for each parameter.

\begin{thm_eigen}[Monotonicity of the class of systems~$\varSigma(\bm p)$]
	\label{thm:boundingSystems}
	Let partial order relations be defined by~\cref{eq:relationOrderVector:y}, \cref{eq:relationOrderVector:xu}, \cref{eq:relationOrderVector:p}, and~$
		\bm r_\varphi =
		\begin{pmatrix}
			- & + & -
		\end{pmatrix}^\top
	$.
	Consider parameter vectors~$
		\bm p^\mathrm{a},
		\bm p^\mathrm{b}
		\in \mathbb{D}_\mathrm{p}
	$
	satisfying ~\cref{eq:performanceSpecification:restriction:c11c12}, initial conditions~$
		\bm x_0^\mathrm{a},
		\bm x_0^\mathrm{b}
		\in \mathbb{D}_\mathrm{x}
	$ and input signals~$
		u^\mathrm{a}(t),
		u^\mathrm{b}(t)
		\in \mathbb{D}_\mathrm{u}
		\forall t \geq 0
	$
	that are ordered according to~\cref{eq:aim:order:parameter,eq:aim:order:initialCondition,eq:aim:order:manipulatedSignal}, respectively.

	Then the corresponding output trajectories~$
		\bm y^\mathrm{a}(t),
		\bm y^\mathrm{b}(t)
	$ and static characteristics~$
		\bm \varphi(\bm p^\mathrm{a}, y_\lambda^\mathrm{eq}),
		\bm \varphi(\bm p^\mathrm{b}, y_\lambda^\mathrm{eq})
	$ of the class of systems~$\varSigma(\bm p)$ are monotone in the sense of~\cref{%
		eq:performanceSpecification:enclose:trajectory,%
		eq:performanceSpecification:enclose:static%
	}.
\end{thm_eigen}
\begin{pf}
	\cref{eq:performanceSpecification:enclose:trajectory} holds, since an increase in the parameter values in the sense of~\cref{eq:relationOrderVector:p} has the same monotone influence on~$\bm x(t)$ and on~$\bm y(t)$
	(Lemmas~\labelcref{%
		lem:processModell:monotone:p:to:x:y,%
		lem:processModell:parameterMonotone%
	})
	as~$\bm x_0$ and~$u(t)$
	(Lemmas~\labelcref{%
		lem:processModell:monotoneOutputFunction:xy,%
		lem:processModell:cooperative%
	}).

	Due to~\cref{eq:performanceSpecification:restriction:c11c12}, the same static reactive gas partial pressure~$x_\mathrm{rg}^\mathrm{eq} = x_\mathrm{rg}^\mathrm{eq,a} = x_\mathrm{rg}^\mathrm{eq,b}$ results for~$\bm p^\mathrm{a}$ and~$\bm p^\mathrm{b}$, which renders~Lemma~\labelcref{lem:processModell:monotone:p:static} applicable such that~\cref{eq:processModel:static:equilibriumPoint:ux:basedOn:xrg} is ordered according to~$\preceq_{\bm r_{\pi}}$.
	Due to Lemmas~\labelcref{lem:processModell:monotoneOutputFunction:xy,lem:processModell:monotone:p:to:x:y},~\cref{eq:processModel:berg2014ss:outputFunction:xtaysb} increases if~$\bm p$ increases with respect to~\cref{eq:relationOrderVector:p} and to~\cref{eq:relationOrderVector:y}, \cref{eq:processModel:static:equilibriumPoint} is monotone with respect to~$\bm r_\varphi$ and~\cref{eq:performanceSpecification:enclose:static} holds true.
\end{pf}

\begin{rem_eigen}
	\label{rem:why_b_needsToBeEqual}
	The exact mapping~\cref{eq:processModel:static:xrg:basedOn:yLambda} from $y_\lambda^\mathrm{eq}$ to~$x_\mathrm{rg}^\mathrm{eq}$, resulting from~\cref{eq:performanceSpecification:restriction:c11c12}, is necessary to obtain the ordered static characteristic~\cref{eq:performanceSpecification:enclose:static}.
	Violation of~\cref{eq:performanceSpecification:restriction:c11c12} leads to different values~$x_\mathrm{rg}^\mathrm{eq,a}$ and~$x_\mathrm{rg}^\mathrm{eq,b}$ for one measured~$y_\lambda^\mathrm{eq}$, and due to the undefined monotonicity of~$x_\mathrm{rg}^\mathrm{eq}$ to~$u^\mathrm{eq}$ (Lemma~\labelcref{lem:analysis:monotone:staticCharacteristic}), it is not possible to determine the order relation for the
	first element of~$\bm r_\varphi$.
	Thus, no intervals for~$c_{11}$ and~$c_{12}$ can be estimated with the method presented in~\cref{sec:intervalEstimation}.
\end{rem_eigen}

\section{Parameter-interval estimation method}
\label{sec:intervalEstimation}

This section addresses objective~(\MakeUppercase{\romannumeral 2}) and presents the new method to determine two parameter vectors~$\bm p^\mathrm{a}$ and~$\bm p^\mathrm{b}$ that, firstly, fulfil the inclusion requirement~\cref{eq:aim:order:parameter} with respect to the parameter vector and, secondly, ensure the inclusion requirements~\cref{eq:performanceSpecification:encloseMeasurement:trajectory,eq:performanceSpecification:encloseMeasurement:static} of the measured steady-state values~$\bm \varphi^\mathrm{M}(y^\mathrm{eq}_\lambda)$ and of the measured output trajectory~$\bm y^\mathrm{M}(t)$.

To this end, the two optimization problems
\begin{alignat}{1}
	\begin{alignedat}{1}
		\hat{\bm p}^\mathrm{a}
		&=
		\argmin_{%
		\bm p^\mathrm{a} \in \mathbb{D}_\mathrm{p}^\mathrm{a}
		}
		J(\bm p^\mathrm{a})
		, \qquad
		\hat{\bm p}^\mathrm{b}
		=
		\argmin_{%
			\bm p^\mathrm{b} \in \mathbb{D}_\mathrm{p}^\mathrm{b}
		}
		J(\bm p^\mathrm{b})
		\\
		&\text{subject to \cref{%
			eq:performanceSpecification:encloseMeasurement:trajectory,%
			eq:performanceSpecification:encloseMeasurement:static%
			}
		}
	\end{alignedat}
	\label{eq:optimizationProblem:bothProblems}
\end{alignat}
have to be solved.
Based on the relative change~$\delta$ of a nominal parameter vector~$\bm p^0$
whose values can be taken from literature and on the constraints in~\cref{eq:performanceSpecification:restriction:c11c12}, a box
\begin{align}
	\mathbb{D}_\mathrm{p}^0
	&=
	\set{
		\bm p
		\in 
		\closedinterval{(1-\delta) \bm p^0}{(1+\delta) \bm p^0}
		\mid
		c_{11} = c_{11}^0
		,
		c_{12} = c_{12}^0
	}
	\label{eq:optimizationProblem:searchDomain:initial}
\end{align}
is defined as the search domain for the parameter vectors.
To guarantee that the parameter values are increased and decreased according to the partial order relation~\cref{eq:aim:order:parameter}, the search domain~$\mathbb{D}_\mathrm{p}^0$ is split into disjunct sets
\begin{align}
	\mathbb{D}_\mathrm{p}^\mathrm{a}
	&=
	\set{
		\bm p
		\in
		\mathbb{D}_\mathrm{p}^0
		\mid
		\bm p \preceq_{\bm r_\mathrm{p}} \bm p^0
	}
	,
	& 
	\mathbb{D}_\mathrm{p}^\mathrm{b}
	&=
	\set{
		\bm p
		\in
		\mathbb{D}_\mathrm{p}^0
		\mid
		\bm p^0 \preceq_{\bm r_\mathrm{p}} \bm p
	}
	.
	\label{eq:optimizationProblem:searchDomain:disjunctSets}
\end{align}

The cost function~$J(\bm p) = J_{\mathrm{y}}(\bm p) + J_{\varphi}(\bm p)$ consists of
\begin{align}
	J_{\varphi}(\bm p)
	&=
	\sum_{q=1}^3
	\sum_{k=1}^{K^\varphi_q}
	\Big(
		\underbrace{
			\varphi^\mathrm{M}_q(y_\lambda^\mathrm{eq}(k))
			-
			\varphi_q(\bm p, y_\lambda^\mathrm{eq}(k))
		}_{\Delta^\varphi_q(k)}
	\Big)^2
	w^\varphi_q(k)
	\label{eq:optimizationProblem:costFunction:static}
	\\
	J_{\mathrm{y}}(\bm p)
	&=
	\sum_{q=1}^2
	\sum_{k=1}^{K^\mathrm{y}_q}
	\Big(
		\underbrace{
			y_q^\mathrm{M}(kT)
			-
			y_q(\bm p, kT)
		}_{\Delta^\mathrm{y}_q(k)}
	\Big)^2
	w^\mathrm{y}_q(k)
	\label{eq:optimizationProblem:costFunction:trajectory}
\end{align}
and corresponds to the element-wise weighted (\cf~$w^\varphi_q(k)$, $w^\mathrm{y}_q(k)$) $l_2$-norm of the differences~$\Delta^\varphi_q(k)$, and $\Delta^\mathrm{y}_q(k)$ between the measured data and the model data.
In~\cref{eq:optimizationProblem:costFunction:trajectory}, $y_q(\bm p, t)$ denotes the solution of~$\varSigma(\bm p)$ for the manipulated signal~$u(t) = u^\mathrm{M}(t)$ used in the experiment and the initial condition~$\bm x_0 = \bm x^\mathrm{eq}$ that is equal to the equilibrium state~(\cf~\cref{eq:processModel:static:equilibriumPoint:ux:basedOn:xrg}) for the initially measured voltage~$y_\lambda^\mathrm{M}(0)$.
$K_q^\varphi$ and~$K_q^\mathrm{y}$ denote the number of measured equilibrium states and the number of sampling points obtained with sampling time~$T$.
The argument~$k$ denotes the respective element in the data series.
The global optimization problems in~\cref{eq:optimizationProblem:bothProblems} are non-convex and thus require much computational time.

Solving~\cref{eq:optimizationProblem:bothProblems} tightens~\cref{eq:optimizationProblem:searchDomain:initial} to the interval defined by~$\hat{\bm p}^\mathrm{a}$ and~$\hat{\bm p}^\mathrm{b}$ and guarantees fulfilment of objective~(\MakeUppercase{\romannumeral 2}) of the parameter-interval estimation method:
the parameter search domains in~\cref{eq:optimizationProblem:searchDomain:disjunctSets} ensure~\cref{eq:aim:order:parameter}, the monotonicity of the static characteristic,~\ie,~$y_\lambda^\mathrm{eq} \mapsto \bm x^\mathrm{eq}$, (\cref{fig:strukturgraph_monotoneStaticFunc_xrg_monotoneOutFunc,fig:strukturgraph_parameterMonotoneStaticFunc_OutFunc}) ensures~\cref{eq:aim:order:initialCondition},
the condition~\cref{eq:aim:order:manipulatedSignal} holds, since~$u^\mathrm{M}(t)$ is used for both models as input and~\cref{eq:performanceSpecification:encloseMeasurement:trajectory,eq:performanceSpecification:encloseMeasurement:static} hold because they are included as constraints in~\cref{eq:optimizationProblem:bothProblems}.

The parameter-interval estimation method is summarized by the following algorithm.

\textbf{Given} are a nominal parameter vector~$\bm p^0$ and measurements of the steady-state values~$\bm \varphi^\mathrm{M}(y^\mathrm{eq}_\lambda)$ and of a trajectory tuple~$(u^\mathrm{M}(t), \bm y^\mathrm{M}(t))$.
\begin{itemize}
	\item[\textbf{1.}] Select a small~$\delta$ and determine~$\bm p^\mathrm{a} = \diag (\bm 1 - \delta \bm r_\mathrm{p}) \bm p^0$ and~$\bm p^\mathrm{b} = \diag (\bm 1 + \delta \bm r_\mathrm{p})\bm p^0$ with the restriction~\cref{eq:performanceSpecification:restriction:c11c12}.
	\item[\textbf{2.}] Verify that~\cref{eq:performanceSpecification:encloseMeasurement:trajectory,eq:performanceSpecification:encloseMeasurement:static} hold true, otherwise increase~$\delta$ and go back to step~1.
	\item[\textbf{3.}] Select positive weights~$w^\varphi_q(k)$, $w^\mathrm{y}_q(k)$ and solve~\cref{eq:optimizationProblem:bothProblems}.
\end{itemize}
The \textbf{result} is two parameter vectors~$\hat{\bm p}^\mathrm{a}$ and~$\hat{\bm p}^\mathrm{b}$ guaranteeing the inclusion properties~\cref{eq:performanceSpecification:encloseMeasurement:trajectory,eq:performanceSpecification:encloseMeasurement:static} and that every parameter vector~\cref{eq:processModel:berg2014ss:parameterVector} satisfying~$\hat{\bm p}^\mathrm{a} \preceq_{\bm r_\mathrm{p}} \bm p \preceq_{\bm r_\mathrm{p}} \hat{\bm p}^\mathrm{b}$ yields a static characteristic~\cref{eq:processModel:static:equilibriumPoint} and output trajectory~$\bm y(t)$ that both are also enclosed in the sense of~\cref{eq:performanceSpecification:encloseMeasurement:trajectory,eq:performanceSpecification:encloseMeasurement:static}.

\section{Experimental validation}
\label{sec:validation}

\begin{figure}[tb]
	\begin{center}
		\includegraphics{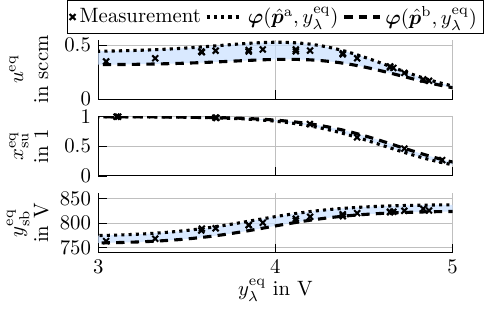}
		\caption{%
			The model output provides an enclosure for the measured steady-state values.
		} 
		\label{fig:fig_identResult_static}
	\end{center}
\end{figure}

\begin{figure}[tb]
	\begin{center}
		\includegraphics{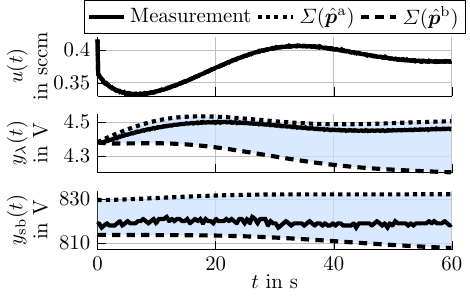}
		\caption{%
		A typical measured trajectory enclosed by the model output defined by two parameter vectors obtained from~\cref{eq:optimizationProblem:bothProblems}.
		} 
		\label{fig:fig_identResult_trajectories}
	\end{center}
\end{figure}

The optimization problems in~\cref{eq:optimizationProblem:bothProblems} are solved, based on measurements of the steady-state values~$\bm \varphi^\mathrm{M}(y^\mathrm{eq}_\lambda)$ and of multiple step responses~($u^\mathrm{M}(t)$, $\bm y^\mathrm{M}(t)$).
$x_\mathrm{su}^\mathrm{eq}$ is measured by energy dispersive X-ray spectroscopy after unloading the coated substrate of the reactor chamber.

\Cref{fig:fig_identResult_static,fig:fig_identResult_trajectories} demonstrate the practical applicability and the correctness of the parameter-interval estimation method, since the measured data is enclosed by the respective model output data for the two parameter vectors~$\hat{\bm p}^\mathrm{a}$ and~$\hat{\bm p}^\mathrm{b}$.
The average relative parameter-interval width~$0.08 = \frac{1}{16} \sum_{i=1}^{16} \frac{\abs{\hat{p}^\mathrm{b}_i - \hat{p}^\mathrm{a}_i} 2}{\abs{\hat{p}^\mathrm{b}_i + \hat{p}^\mathrm{a}_i}}$ shows that the estimation method is able to determine a tight parameter interval for the wide range of~$y_\lambda^\mathrm{eq}$ depicted in~\cref{fig:fig_identResult_static}.
Due to Theorem~\labelcref{thm:boundingSystems}, every parameter vector~$\bm p$ satisfying~$
	\hat{\bm p}^\mathrm{a}
	\preceq_{\bm r_\mathrm{p}}
	\bm p
	\preceq_{\bm r_\mathrm{p}}
	\hat{\bm p}^\mathrm{b}
$
leads to static characteristics and output trajectories that are enclosed by the interval depicted in~\cref{fig:fig_identResult_static,fig:fig_identResult_trajectories}.

If additional equilibrium points with~$y_\lambda^\mathrm{eq} < \qty{3}{\volt}$, or output trajectories spanning more than~\qty{0,5}{\volt}, are taken into account, excessively large parameter intervals must be allowed by~$\delta$ (\cf~\cref{eq:optimizationProblem:searchDomain:initial}) and result from~\cref{eq:optimizationProblem:bothProblems}.
This suggests that the model's structural validity is confined to regions around specific operating points.

\section{Conclusion}
\label{sec:conclusion}

This contribution classifies the well-established process model of reactive sputtering as cooperative and parameter-monotone.
Based on this result, a novel parameter-interval estimation method is introduced that provides a guaranteed enclosure of the measurements.
Experimental results demonstrate its correctness and practical applicability.
The method enables the determination of a parameter set that allows the design of an interval observer contributing to the monitoring task for reactive sputtering.

Future work will investigate more sophisticated process models and generally informative plasma variables, aiming, \eg, to further reduce the parameter interval width.

\begin{ack}
	The authors gratefully acknowledge financial support from the Deutsche Forschungsgemeinschaft (DFG) through grant 562363182.
\end{ack}

\bibliography{2026_ifacconf}             

\begin{thebibliography}{12}
\providecommand{\natexlab}[1]{#1}
\providecommand{\url}[1]{\texttt{#1}}
\providecommand{\urlprefix}{URL }
\expandafter\ifx\csname urlstyle\endcsname\relax
  \providecommand{\doi}[1]{doi:\discretionary{}{}{}#1}\else
  \providecommand{\doi}{doi:\discretionary{}{}{}\begingroup
  \urlstyle{rm}\Url}\fi

\bibitem[{Angeli and Sontag(2003)}]{Angeli:2003}
Angeli, D. and Sontag, E.D. (2003).
\newblock Monotone control systems.
\newblock \emph{IEEE Trans. Autom. Control}, 48(10), 1684--1698.

\bibitem[{Berg and Nyberg(2005)}]{Berg:2005}
Berg, S. and Nyberg, T. (2005).
\newblock Fundamental understanding and modeling of reactive sputtering
  processes.
\newblock \emph{Thin Solid Films}, 476(2), 215--230.

\bibitem[{Berg et~al.(2014)Berg, Särhammar, and Nyberg}]{Berg:2014}
Berg, S., Särhammar, E., and Nyberg, T. (2014).
\newblock Upgrading the “{B}erg-model” for reactive sputtering processes.
\newblock \emph{Thin Solid Films}, 565, 186--192.

\bibitem[{Jaulin et~al.(2001)Jaulin, Kieffer, Didrit, and Walter}]{Jaulin:2001}
Jaulin, L., Kieffer, M., Didrit, O., and Walter, {\'E}. (2001).
\newblock \emph{Applied Interval Analysis}.
\newblock Springer, London.

\bibitem[{Kelemen and Madarász(2021)}]{Kelemen:2021}
Kelemen, A. and Madarász, R.R. (2021).
\newblock Reactive magnetron sputtering: an offline parameter identification
  method.
\newblock In \emph{Proc. of the IEEE Int. Symp. on Applied Computational
  Intelligence and Informatics}, 357--362.

\bibitem[{Kieffer and Walter(2011)}]{Kieffer:2011}
Kieffer, M. and Walter, E. (2011).
\newblock Guaranteed estimation of the parameters of nonlinear continuous-time
  models: Contributions of interval analysis.
\newblock \emph{Int. J. Adapt. Control Signal Process.}, 25(3), 191--207.

\bibitem[{Kubart et~al.(2006)Kubart, Kappertz, Nyberg, and Berg}]{Kubart:2006}
Kubart, T., Kappertz, O., Nyberg, T., and Berg, S. (2006).
\newblock Dynamic behaviour of the reactive sputtering process.
\newblock \emph{Thin Solid Films}, 515(2), 421--424.

\bibitem[{Mahne et~al.(2022)Mahne, Čekada, and Panjan}]{Mahne:2022}
Mahne, N., Čekada, M., and Panjan, M. (2022).
\newblock Total and differential sputtering yields explored by {SRIM}
  simulations.
\newblock \emph{Coatings}, 12(10), 1541.

\bibitem[{Rudin(1976)}]{Rudin:1976}
Rudin, W. (1976).
\newblock \emph{Principles of Mathematical Analysis}.
\newblock McGraw-Hill, New York, 3rd edition.

\bibitem[{Schneider and Woelfel(2023)}]{Schneider:2023:ccta}
Schneider, F. and Woelfel, C. (2023).
\newblock Robust interval observer for substrate state estimation of nonlinear
  reactive sputter processes.
\newblock In \emph{Proc. of the IEEE Conf. on Control Technology and
  Applications}, 194--200.

\bibitem[{Smith(1995)}]{Smith:1995}
Smith, H.L. (1995).
\newblock \emph{Monotone Dynamical Systems}.
\newblock American Mathematical Society, Providence.

\bibitem[{Strijckmans et~al.(2012)Strijckmans, Leroy, {De Gryse}, and
  Depla}]{StrijckmansLeroyGryseDepla:2012}
Strijckmans, K., Leroy, W.P., {De Gryse}, R., and Depla, D. (2012).
\newblock Modeling reactive magnetron sputtering: Fixing the parameter set.
\newblock \emph{Surf. Coat. Technol.}, 206(17), 3666--3675.

\end{thebibliography}

\end{document}